# Risk-limiting Load Restoration for Resilience Enhancement with Intermittent Energy Resources

Zhiwen Wang, *Student Member, IEEE,* Chen Shen, *Senior Member, IEEE,* Yin Xu, *Member, IEEE,* Feng Liu, *Member, IEEE,* Xiangyu Wu, *Student Member, IEEE,* Chen-Ching Liu, *Fellow, IEEE*

*Abstract*—Microgrids are resources that can be used to restore critical loads after a natural disaster, enhancing resilience of a distribution network. To deal with the stochastic nature of intermittent energy resources, such as wind turbines (WTs) and photovoltaics (PVs), many methods rely on forecast information. However, some microgrids may not be equipped with power forecasting tools. To fill this gap, a risk-limiting strategy based on measurements is proposed. Gaussian mixture model (GMM) is used to represent a prior joint probability density function (PDF) of power outputs of WTs and PVs over multiple periods. As time rolls forward, the distribution of WT/PV generation is updated based the latest measurement data in a recursive manner. The updated distribution is used as an input for the risk-limiting load restoration problem, enabling an equivalent transformation of the original chance constrained problem into a mixed integer linear programming (MILP). Simulation cases on a distribution system with three microgrids demonstrate the effectiveness of the proposed method. Results also indicate that networked microgrids have better uncertainty management capabilities than stand-alone microgrids.

## Nomenclature

Most symbols used in this paper are listed below for ease of reference; others are explained following their first appearance.

**Indices**

| | |
|---|---|
| $k, t$ | Indices for periods. And, $t=k+1,\ldots,K$ |
| $g$ | Indices for generators and energy storage systems (ESSs) |
| $v$ | Indices for WTs and PVs |
| $l$ | Indices for loads |
| $m$ | Indices for Gaussian components |

**Parameters**

| | |
|---|---|
| $K$ | Number of periods |
| $G$ | Number of generators and ESSs |
| $W$ | Number of WTs |
| $S$ | Number of PVs |
| $L$ | Number of loads |
| $M$ | Number of Gaussian components |
| $\omega_m, \boldsymbol{\mu}_m, \boldsymbol{\Sigma}_m$ | Parameters of the $m$th component of a GMM |
| $\Omega$ | $\Omega=\{\omega_m, \boldsymbol{\mu}_m, \boldsymbol{\Sigma}_m \mid m=1,\ldots,M\}$ |
| $p_g^{\max}, p_g^{\min}$ | Maximum and minimum power outputs of generator $g$ |
| $c_l$ | Priority weight of load $l$ |
| $\tau$ | Length of each period. $\tau=1$ hour |
| $r_g^{\text{up}}, r_g^{\text{dn}}$ | Ramp-up and ramp-down limits of generator $g$ |
| $p_{t,l}$ | Power of load $l$ in period $t$ |
| $EN_g$ | Generation resource of generator $g$ |
| $EC_g$ | Capacity of ESS $g$ |
| $\rho_c, \rho_d$ | Charging and discharging efficiency of ESS $g$ |

**Random Variables**

| | |
|---|---|
| $X_t^v$ | power output of WT/PV $v$ in period $t$ |

**Decision Variables**

| | |
|---|---|
| $p_{t,g}$ | Scheduled power output of generator $g$ in period $t$ |
| $P_{t,g}^{\text{ch}}, P_{t,g}^{\text{dch}}$ | Scheduled charging/discharging power of ESS $g$ in period $t$ |
| $u_{t,l}$ | Status of load $l$ to determine in period $t$. $u_{t,l} = 1$ if load $l$ restored in period $t$; $u_{t,l} = 0$, otherwise |

## I. Introduction

### A. Motivation

Extreme weather events, such as flooding and hurricanes, cause severe damages to power systems from time to time. For example, hurricane Sandy hit the east coast of U.S. in 2012, causing power service interruption for 7.5 million people [1]. In 2008, a snow storm in Southern China left over 14 million households out of power for days. To help power systems defend against major disasters, the concept of "resilience" is proposed. Resilience refers to the ability of a power system that can withstand low-probability high-impact events and quickly recover customer service from major disasters [2]-[4]. Distributed energy resources (DERs), energy storage systems (ESSs), as well as microgrids (MGs), are valuable resources that can enhance resiliency of a distribution system, as they can be used to restore critical loads when the utility power is not available.

### B. Literature Review

Much research has been conducted on resilience-oriented load restoration problems [5]-[7]. A three-stage restoration method for distribution networks is proposed in [5] that can be used to maximize the restored loads with available DERs. Authors in [6] study an automatic reconfiguration algorithm of microgrids with dynamical boundaries, considering transients of line faults and generators. In [7], a method is proposed to transform a load restoration problem into a linear integer programming "*maximum coverage*," resulting in an efficient solution. However, these studies do not fully take renewable DERs into consideration. One prominent feature of renewables is the stochastic nature of generation. If uncertainties are not fully considered in the load restoration procedure, undesirable events, e.g., load shedding, over/under-frequency issues, may happen. To reduce these risks, there is a great need for a systematic method to manage uncertainties introduced by renewable DERs.

There are several popular methodologies in the literature that address the renewable uncertainty issue. The first approach is using a forecast value to replace the actual wind/solar power. For example, authors in [8] propose an operation and self-healing strategy, regarding the power outputs of DERs as deterministic forecast values. In a relevant study [9], based on short-term forecasts, a model predictive control approach is proposed to enhance power system resilience with microgrids. These studies only consider forecasted scenarios. However, as the actual values of renewable generation may deviate far from the forecasts, these methods may not be appropriate.

The second approach is called "scenario-based stochastic optimization". This method generates many possible scenarios with different probabilities, and determines an optimal strategy that can minimize an expected cost in those scenarios. In [10], forecast errors of wind/solar power are modeled by *Beta* distributions. By running Monte Carlo simulation, one can generate a number of scenarios of DG outputs. To reduce the computational burden, the back forward reduction method is adopted to retain representative scenarios. In a relevant study [11], a two-stage stochastic programming model is proposed, in which uncertainty scenarios are generated by auto-regressive-moving-average (ARMA) technique based on weather forecast information. Although the scenario-based stochastic optimization is able to consider different possibilities of uncertainties, there is a drawback. In order to keep the information of uncertainties as intact as possible, the number of scenarios has to be large. As a consequence, the computation of the approach becomes intractable.

The third approach is robust optimization (RO). In [12], robust optimization is proposed to handle renewable and load uncertainties, in which uncertain parameters are modeled by convex and bounded intervals, e.g., $\pm 20\%$ of installed capacities. In [13], a two-stage robust optimization model is proposed to coordinate the hardening and DER allocation. Based on weather forecast, the spatial and temporal dynamics of an uncertain natural disaster, e.g., a hurricane, are captured by a multi-stage and multi-zone uncertainty set. Because RO can be transformed into a single-level *min* (or *max*) problem, the computation of RO is relatively inexpensive compared with the scenario-based approach. However, as the optimal solution of RO is always obtained at an extreme point, i.e., the worst-case scenario, RO suffers from the inherent conservativeness, which decreases the overall efficiency of an optimal strategy.

In recent years, an important methodology that is developed to deal with uncertainties, called "risk-limiting dispatch (RLD)", has drawn more and more attention [14]-[19]. RLD aims at finding an optimal strategy with a minimum cost to restrain risks brought by uncertainties, e.g., the insufficient generation risk. In RLD, power delivered at a particular period is determined in multiple decision stages. In each decision stage, distributions of renewable uncertainties are continuously updated as conditional distributions of the latest forecasts, e.g., the forecasted wind speed from a numerical weather prediction (NWP) system, wind power forecast values from a forecasting tool. Thereafter, optimal decisions are made to limit risks based on the newly collected information, i.e., decision variables depend on the prediction information. If more precise information of uncertainties is obtained in future decision stages, operators have a chance to take recourse actions to remedy previous decisions. At the moment the uncertainty is realized, the risk is restrained within a tolerance. As the accuracy of descriptions of uncertainties increases with newly collected information, it is less conservative. The original idea of RLD is proposed in [14]. In [15], an analytical solution of a simple RLD is obtained based on dynamic programming principles. The congestion issue is discussed in [16], while the wind power ramping issue is addressed in [17]. The applications of RLD in unit commitment and economical dispatch can be found in [18], [19]. The scope of this paper is a load restoration problem with microgrids in the distribution system level. A major concern on the application of RLD in the distribution level is the absence of power forecasting tools, which may result in infeasibility of the forecast-based RLD.

Existing methods can be categorized into two groups:

(1) The first one assumes that there is continually updated forecast information, e.g., [8]-[11], [13]-[19].

(2) The second one assumes that uncertainty models are fixed. For example, in [12], the uncertain parameters belong to a fixed interval [-0.2 +0.2], according to personal experience or historical data analysis. In a relevant study [21], renewable uncertainties are modeled by fixed distributions.

In fact, both assumptions might be invalid: (1) in the distribution level, forecasting tools may not be available; (2) In a multiple-period optimization problem, uncertainty models of WTs/PVs in different periods are not the same. Using fixed models may result in conservative results.

In addition to the challenge arising from the limited forecast information, there is one more challenge for RLD: the correlation of renewables. In RLD, probabilistic constraints associated with risks require arithmetical addition operations of random variables, e.g., the power balance equation. When random variables are correlated, "convolution technique" is no longer applicable [20]. Therefore, it is difficult to transform the probabilistic constraints into tractable forms.

*C. Contributions*

To address these important issues, this paper proposes a risk-limiting load restoration strategy based on observations of renewable uncertainties. Relative to the state-of-the-art, the

contributions of this paper are threefold:

(1) The resilience-oriented customer service restoration with intermittency renewables is formulated as a risk-limiting decision-making problem.

(2) In modeling of uncertainties, a methodology is proposed to recursively update distributions of renewables as latest observations of WT/PV generation are available. Both wind and solar power uncertainties are modeled in a universal manner. The spatial-temporal correlation of uncertainties is considered.

(3) For the solution methodology, probabilistic constraints consisting of correlated random variables are converted into linear inequalities, based on a property of the updated distribution. By doing so, the original problem can be solved with commercial solvers.

*D. Organization*

The remainder of the paper is organized as follows: Section II provides the problem formulation; Section III discusses a method to update distributions of uncertainties based on observations. In Section IV, the original problem is transformed into an MILP. Case study results are presented in Section V. Conclusion and limitations are provided at the end of this paper.

## II. PROBLEM FORMULATION

This section provides the formulation of the risk-limiting load restoration problem. First, assumptions are provided and justified. Then, the resilience metric used in this paper is elaborated. A framework is given to clarify the computational procedure, followed by a full description of the constraints of the problem. Finally, challenges of the problem are discussed.

*A. Assumptions*

(1) There is a centralized microgrid controller that collects information and optimizes the operation. Intelligent electronic devices (IEDs) receive data and measurements from sensors and pass them to the central controller. IEDs also transfer control commands from the central controller to microgrid devices, such as diesel generators, ESSs, and load breakers. This communication structure can be implemented following the standard IEC 61850 via TCP/IP [22]. In fact, it is commonly used in microgrid testbeds [23], [24].

(2) In order to focus on volatile renewables, this paper assumes that uncertainties come from solar and wind generation only. Load uncertainty is not considered. This is justified as follows. Note that only a small number of critical loads will be restored after a major disaster. Since the critical loads are used to maintain basic and minimum societal functions, e.g., hospitals and street lighting, their demand does not change significantly during the restoration process [25]. Therefore, load variations are much smaller than fluctuations of renewable generation. Thus, they are neglected. In addition, if load varies during the restoration process, the maximum load demand will be used as the input to the restoration problem [25], [26], which ensures that the load demand will be satisfied.

*B. Resilience Metric*

The conceptual resilience curve in Fig. 1 that is associated with an extreme event is proposed in [27], and adopted in [25], [28]-[30]. There are 7 well-defined states, in which different measures can be taken to improve the resilience level, e.g., resource preparation in the pre-event state, and service restoration using microgrids in the restorative state and post-restoration state.

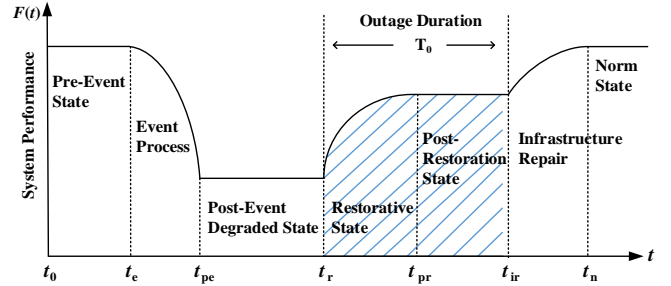

Fig. 1 A conceptual resilience curve

This paper is concerned with a load restoration strategy to enhance resilience. Therefore, the period from $t_r$ to $t_{ir}$ is of primary interest, i.e., the restorative state and post-restoration state. When an extreme event takes place, the utility power is not available from $t_r$ to $t_{ir}$. Hence, the concerned period is referred to as the "outage duration". The resilience level is evaluated by the integral of the system performance function over the outage duration [27]-[30], i.e., the blue-shadowed area in Fig.1:

$$R = \int_{t_r}^{t_{ir}} F(t) dt \quad (1)$$

The system performance function adopted in this paper is similar to those used in relevant resilience studies [25], [26], [28], [29]: the weights of restored loads. This is reasonable because more critical loads being restored indicates that the distribution system can better restore service to customers after an extreme event, i.e., the system is more resilient. Priority weights of critical loads (hospital, police station, etc.,) should be sufficiently greater than that of non-critical ones so that critical loads can be served as a priority with limited available generation capacities and resources.

Let the outage duration $T_0$ be discretized into several sub-intervals τ. Thereafter, resilience index (1) can be computed as follows:

$$R = \sum_t \sum_l c_l \cdot u_{t,l} \cdot \tau \quad (2)$$

That is to say, the resilience level is evaluated by the cumulative service time to loads weighted by their priority over the outage duration.

**Remark 1:** In some studies [26], [28], the active power of a load is involved in the resilience metric. Then, equation (2) becomes:

$$R = \sum_t \sum_l c_l \cdot P_{t,l} \cdot u_{t,l} \cdot \tau \quad (3)$$

Both (2) and (3) can well represent resilience of a system. In fact, if one regards the *weighted power* in (3) as the *weight* in (2), metrics (2) and (3) are equivalent. Regarding this, both metrics can be applied to the proposed risk-limiting load restoration method. This paper prefers metric (2) because it is more intuitive than (3). For example, a hospital should be identified as a critical load with a large weighted coefficient, as it is a life-critical infrastructure after an extreme event. Suppose that the hospital consumes a small amount of active power. If metric (3) is adopted, the hospital might not be restored due to a small value of the weighted power, while metric (2) avoids the problem.

**Remark 2:** It should be noted that the resilience index is quite different from the traditional "reliability" concept.

Reliability is associated with common events, most of them making a low impact, e.g., a short circuit on a line. For such events, most power sources stay connected. Transmission lines and distribution feeders are available. Since the power system facilities are intact, several scenarios will lead to high reliability, e.g., scheduling sufficient reserves in preparation for any single failure of equipment, and utilization of automatic reclosers to re-energize a feeder after a temporary fault. By doing so, reliability indices, e.g., the system average interruption frequency index (SAIFI) and system average interruption duration index (SAIDI), will be low.

In contrast, resilience is designed for extreme events that rarely happen (low-probability high-impact events), during which the power system infrastructure, e.g., transmission lines, substations, automatic reclosers, and remote-control switches, might be damaged. As a result, numerous power system components are not available. Traditional measures that are designed for reliability enhancement may not apply. For example, if the upstream substation is damaged by a hurricane, the automatic recloser cannot re-energize downstream loads. To improve the resilience level, i.e., to quickly recover from a major disaster, focus should be placed on how to pick up critical loads as many as possible when the utility power is unavailable. Measures for the resilience enhancement include utilization of microgrids to restore critical service and hardening of transmission and distribution facilities.

More discussions on the differences between resilience and reliability can be found in [31].

*C. Framework*

RLD designed for transmission systems is based on two assumptions. First, as time moves forward, the sensing system can obtain more precise information on uncertainties. Second, there is a market mechanism, where a power/energy transaction enforced in earlier times is less costly. Therefore, system operators using RLD scheme determine the power/energy delivered at a certain period in multiple decision stages, trying to achieve a tradeoff between making decisions earlier with cheaper prices and making decisions later with more precise information. In each decision stage, based on the latest information, the risks brought by uncertainties are limited within a tolerance, i.e., constraints associated with risk indices are added into optimization models.

For a load restoration problem, the first assumption of RLD still holds, while the second one may not be valid. An example is given in Fig. 2. For the delivery period between 9:00 to 10:00, suppose that one can make decisions in three stages, 7:00, 8:00, and 9:00, which are called the 1$^{st}$, 2$^{nd}$, and final decisions, respectively. Clearly, the uncertainty information of the delivery period obtained at 9:00 is more precise than that obtained at 8:00, which is more precise than that obtained at 7:00. However, unlike the market mechanism in the transmission system level, the weighted coefficients of resilience index (2) are independent of time, i.e., enforcing decisions that are made earlier than 9:00 will not bring additional reward. Therefore, a rational decision maker will not enforce the decisions made at 7:00 and 8:00. That is to say, the 1$^{st}$ and 2$^{nd}$ decisions are for advisory purpose only; they are not implemented.

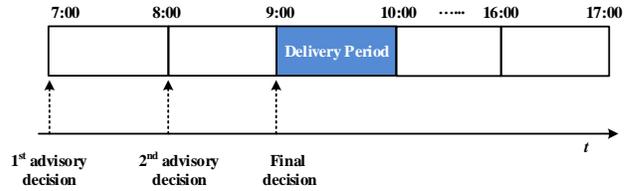

Fig. 2 Illustration of advisory decisions and final ones for a delivery period.

The proposed risk-limiting load restoration is designed as follows. The framework is illustrated in Fig. 3. For demonstration, the time resolution is set to be 1 hour in this paper. It can be changed to other frames, e.g. 10 min or 30 min, if needed. The proposed method is an online tool, which is executed periodically. In each implementation, the proposed method aims to maximize resilience index (2), and determines the set of loads to be restored, the generation schedule of dispatchable generators, and the charging/discharging strategy of ESSs. An optimal load restoration strategy is determined over the optimization window, in which only decisions in the first time interval are implemented to instruct generators/ESSs/loads for the next period $k+1$ whereas the rest are abandoned. There are multiple periods within a certain optimization window. In each period, constraints associated with risks are considered based on the latest uncertainty information. Specifically, this paper uses observations of uncertainties in the past periods to update distributions of renewable generation in the future.

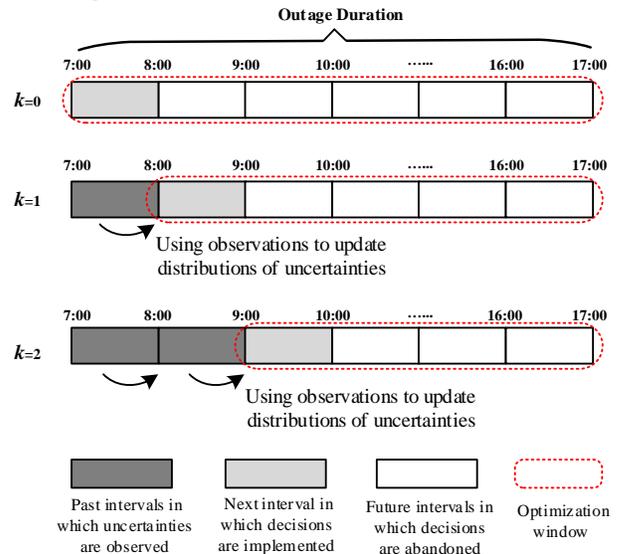

Fig. 3 A framework of the risk-limiting load restoration scheme.

**Remark 3:** The periodic process in Fig. 3 is also known as receding horizon control [32]. Compared with the method in [32], the proposed method has two major differences:

(1) The method in [32] addresses critical issues of a load restoration problem in deterministic scenarios. Renewable uncertainties have not been fully considered [32]. This paper mainly aims at handling uncertainties introduced by renewables in the load restoration process.

(2) Authors in [32] point out that the information should be updated periodically. One original contribution of this paper is a recursive methodology in Section III that uses observations to update distributions of renewable uncertainties.

*D. Constraints*

*1) Risk-limiting constraints*

- Since power outputs of WTs/PVs are stochastic, the following adequacy requirement should be met in each period:

$$\Pr\left\{\sum_{g=1}^{G} P_{t,g} + \sum_{v=1}^{W+S} X_t^v \geq \sum_{l=1}^{L} u_{t,l} P_{t,l} \right\} \geq \alpha \quad \forall t \geq k+1 \quad (4)$$

Equation (4) indicates that the probability of restored load demands being satisfied with available resources should be greater than a given confidence level. This is a typical risk-limiting constraint [13]. In this paper, $\alpha$ is 90%.

- Similar to (4), an energy adequacy requirement is formulated as follows:

$$\Pr\left\{\begin{array}{c}\sum_{g \notin ESS} EN_g(\kappa+1) + \sum_{t=k+1}^{T}\sum_{v=1}^{W+S} X_t^v \tau \\ \geq \sum_{t=k+1}^{T}\sum_{l=1}^{L} u_{t,l} P_{t,l} \tau\end{array}\right\} \geq \alpha \quad (5)$$

$$EN_g(k+1) = EN_g(0) - \sum_{\bar{t}=1}^{k} \bar{P}_{\bar{t},g} \quad \forall g \notin ESS \quad (6)$$

where $EN_g(0)$ is generation resource available at the beginning; $\bar{P}_{\bar{t},g}$ is the power output of generator $g$ in the past $k$ periods.

Note that when real time $k$ rolls forward, distributions of random variables $X_t^v$ in (4)(5) should be updated, which are conditioned on the latest information.

*2) Deterministic constraints*

- If $g$ is a dispatchable generator, e.g., a diesel, then the following output limit, generation resource limit, and ramping limit should be satisfied:

$$P_g^{\min} \leq P_{t,g} \leq P_g^{\max} \quad \forall t \geq k+1, \, g \notin ESS \quad (7)$$

$$\sum_{t=k+1}^{T} P_{t,g} \tau \leq EN_g(k+1) \quad \forall g \notin ESS \quad (8)$$

$$r_g^{dn} \leq P_{t,g} - P_{t-1,g} \leq r_g^{up} \quad \forall t \geq k+2, \, g \notin ESS \quad (9)$$

- If $g$ is an ESS, then two auxiliary binary variables are introduced: $\chi_{t,g}$ being 1 represents the discharging mode; while $\gamma_{t,g}$ being 1 for charging mode. Note that an ESS cannot be in charging and discharging modes at the same time [8]. That is :

$$\chi_{t,g} + \gamma_{t,g} \leq 1 \quad \forall t \geq k+1, \, g \in ESS \quad (10)$$

Besides, limits on charging/discharging power output and maximum/minimum *SOC* are formulated as follows:

$$0 \leq P_{t,g}^{dch} \leq \chi_{t,g} P_g^{dch,\max} \quad, \quad \forall t \geq k+1, \, g \in ESS \quad (11)$$

$$-\gamma_{t,g} P_g^{ch,\max} \leq P_{t,g}^{ch} \leq 0 \quad, \quad \forall t \geq k+1, \, g \in ESS \quad (12)$$

$$SOC_g^{\min} \leq SOC_{t,g} \leq SOC_g^{\max} \quad (13)$$
$$\forall t \geq k+1, \, g \in ESS$$

Note that the use of an ESS would involve decisions in multiple periods, as the current *SOC* depends on the previous *SOC* and the recent charging/discharging strategy. So, the *SOC* transition of an ESS is formulated as follows:

$$SOC_{t,g} = SOC_{t-1,g} - \tau\left(P_{t-1,g}^{dch}\rho_d^{-1} + P_{t-1,g}^{ch}\rho_c\right)/EC_g \quad (14)$$
$$\forall t \geq k+2, \, g \in ESS$$

In the literature [8], the *SOC* at the end is required to be the initial value. This paper does not include such a constraint, because there are necessities to charge an ESS after the outage.

*E. Challenges*

To solve the problem (2)-(14), two critical issues need to be addressed:

(1) Updating distribution of $X_t^v$ as time rolls forward;

(2) Dealing with probabilistic constraints (4) and (5), e.g. computing the distribution of an aggregation of $X_t^v$ from multiple WTs/PVs in a single period, as well as over the whole outage duration. Usually, one may use the "convolution technique" [20] to compute the summation of $X_t^v$. However, $X_t^v$ is interdependent. Hence, the "convolution technique" does not apply.

The two issues are addressed in Section III and IV, respectively.

### III. MODELING UNCERTAINTY

In this section, features of intermittent energy resource uncertainties are discussed, followed by details for modeling, as well as updating, the uncertainties of WT/PV generation.

*A. Features of uncertainties in microgrids*

Wind and solar power are commonly used in microgrids. Their uncertainties have three features:

(1) A distribution of wind power is different from that of solar power. Usually, wind speed follows a *Weibull* distribution [33]. Thereafter, the distribution of wind power can be obtained by the "wind speed-power curve", which is a modified *Weibull* distribution with two spikes at 0.0 and 1.0 p.u. The PV generation production is usually a function of solar irradiance and temperature. Authors in [34] report that the irradiance in Arizona can be described reasonably by a *Beta* distribution, and the temperature is *Gaussian*. Then, the distribution of solar power is obtained as a variant of a *Beta* distribution. Alternatively, authors in [35] adopt the *Saunier* model to evaluate the irradiation and obtain a different distribution of solar generation. In the presence of both WTs and PVs, different forms of distributions make it difficult to solve a load restoration problem in a universal manner.

(2) Wind/solar power outputs have spatial-temporal correlations. First, in the distribution system level, renewables are geographically closely located. Hence, the spatial dependence of power outputs from adjacent wind farms/solar arrays is significant. Second, power outputs over adjacent periods are also correlated. An illustration of the spatial-temporal correlation of two adjacent WTs over 10 hours is shown in Fig. 4. The hourly data of wind power is from the "*wind integration data set*" of National Renewable Energy Laboratory (NREL) [36]. Different colors correspond to different values of correlation coefficients. According to Fig. 4, the correlation coefficients of the two WTs in one period (spatial correlation), and two adjacent periods of one WT (temporal correlation), are about 0.89-0.96, indicating strong interdependences. Similar results can be obtained for PVs.

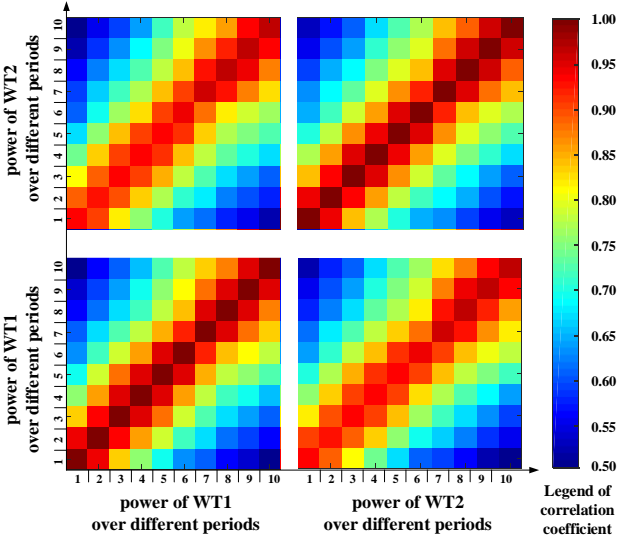

Fig. 4 The spatial-temporal correlation matrix of two WTs over 10 hours

(3) Unlike transmission systems, some microgrids may not be equipped with power forecasting tools due to cost considerations. Hence, in some cases, forecast information for WTs/PVs is unavailable, increasing the difficulty in decision making of load restoration.

To overcome this hurdle, an intuitive idea is to use latest observations to infer future uncertainties. Note that power outputs in period $k$ and $k+1$ are usually positively correlated. If the power in $k$ is observed and it is at a high value, there is a good chance that power in $k+1$ is also at a high value. By doing so, operators can update the inference and make decisions correspondingly for the next periods. Based on this idea, one classic approach, called "persistence forecasting", is developed [37], which uses past measurements of renewable power outputs as forecasted values for the next periods. "Persistence forecasting" is low cost. However, it only obtains an estimated point, not a distribution, of future uncertainties. Therefore, the classic "persistence forecasting" is not suitable for risk-limiting constraints (4)(5). If one wants to use persistence forecasting in the load restoration problem, the probabilistic constraints (4) (5) need modifications. A detailed discussion is provided in Appendix D.

In the following, a method that uses observations to infer distributions of renewable generation for the next periods is detailed.

*B. A prior distribution of uncertainties*

Let a random variable $X_k$ denote the actual power output of an intermittent energy resource in period $k$. Then, a random vector, $X$, for outputs over $K$ periods is defined as:

$$\boldsymbol{X} = [X_1 \ \cdots \ X_k \ \cdots \ X_K]^\mathrm{T} \quad (15)$$

In this paper, a prior joint distribution of $X$ is presented by a GMM with an adjustable parameter set $\Omega=\{\omega_m, \boldsymbol{\mu}_m, \boldsymbol{\sigma}_m\ ; m=1,\ldots,M\}$ as follows:

$$f_X(\boldsymbol{x}) = \sum_{m=1}^{M} \omega_m N_m(\boldsymbol{x}\ ;\ \boldsymbol{\mu}_m, \boldsymbol{\sigma}_m) \quad (16)$$

$$\sum_{m=1}^{M} \omega_m = 1, \quad \omega_m > 0 \quad (17)$$

$$N_m(\boldsymbol{x}\ ;\ \boldsymbol{\mu}_m, \boldsymbol{\sigma}_m) := \frac{e^{-\frac{1}{2}(\boldsymbol{x}-\boldsymbol{\mu}_m)^\mathrm{T} \boldsymbol{\sigma}_m^{-1}(\boldsymbol{x}-\boldsymbol{\mu}_m)}}{(2\pi)^K \det(\boldsymbol{\sigma}_m)^{1/2}} \quad (18)$$

where $\omega_m$ is the weight, $N_m(\cdot)$ is the $m$th multivariate Gaussian component with a mean vector $\boldsymbol{\mu}_m$ and a covariance matrix $\boldsymbol{\sigma}_m$.

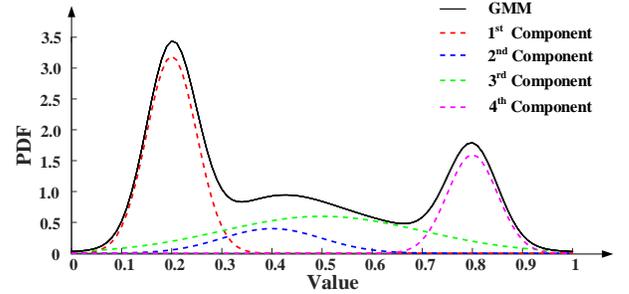

Fig. 5 An illustrative example using a GMM to represent the PDF of a random variable. In this example, $K=1$ and $M=4$. Weights for components are 0.4, 0.1, 0.3, and 0.2, respectively.

In Fig. 5, the idea of using a GMM to represent the PDF of a random variable is presented. For illustration, suppose that the number of periods $K=1$ and the number of Gaussian components $M=4$. The PDF of a GMM (in black) is a weighted summation of four Gaussian distributions (in red, blue, green, and purple, respectively). By adjusting the parameter set $\Omega$, GMM is able to characterize different kinds of non-Gaussian correlated random variables [38]-[40]. Thus, it is suitable for modeling uncertainties of WTs and PVs.

With historical data of $X$, the parameter set $\Omega$ of a GMM can be determined through the maximum likelihood estimation technique [41]. Specifically, this paper adopts a solver *gmdistribution.fit* in MATLAB to estimate the parameter set $\Omega$.

*C. Using observations to update the prior distribution*

Suppose in the second period, the actual power output in the first period is observed. That is,

$$X_1 = x_1 \quad (19)$$

Then, the distribution of the remaining entries, $X_{2,\ldots,K}$, can be updated as a conditional distribution with respect to $X_1=x_1$. In Appendix A, the computational procedure of a conditional distribution is provided:

$$f_{X_{2,\cdots,T}|X_1}(x_{2,\cdots,K}/x_1) \tag{20}$$
$$= \sum_{l=1}^{M} \omega_l'(x_1, \Omega) N_l(x_{2,\cdots,K}; \mu_l(x_1, \Omega), \sigma_l(x_1, \Omega))$$

Note that the updated distribution of $X_{2,\ldots,K}$ has a GMM form. This is an important property of GMM, namely, *conditional invariance*. When the third period comes with an observed realization of $X_2$, the distribution of $X_{3,\ldots,K}$ can be updated in a similar way as (20) via the formulae in Appendix A. A more general updating formula for the period $k+1$ based on an observation of $X_k$ is given as follows:

$$f_{X_{k+1,\cdots,K}|X_{1,\cdots,k}}(x_{k+1,\cdots,K}/x_{1,\cdots,k}) = \sum_{l=1}^{M} \omega_l'(x_{1,\cdots,k}, \Omega) \tag{21}$$
$$\times N_l(x_{k+1,\cdots,K}; \mu_l(x_{1,\cdots,k}, \Omega), \sigma_l(x_{1,\cdots,k}, \Omega))$$

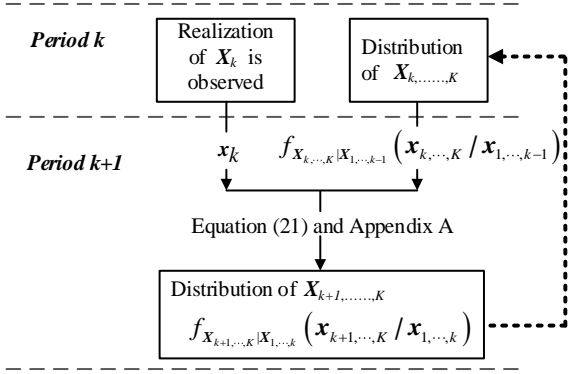

Fig. 6 A recursive procedure to update the distribution of power uncertainties

A flow chart is given in Fig. 6. The distribution of power uncertainties for the next periods with respect to observations in the past can be updated in a recursive manner.

*D. Multiple renewable case*

If there are multiple PVs and WTs, the random vector $X$ is augmented as follows:

$$X = [X_1^1 \cdots X_1^{W+S} \cdots X_k^1 \cdots X_k^{W+S} \cdots X_K^1 \cdots X_K^{W+S}]^T \tag{22}$$

In period $k$, an aggregated observation of $W+S$ components is obtained, i.e.,

$$[X_k^1 \cdots X_k^{W+S}] = [x_k^1 \cdots x_k^{W+S}] \tag{23}$$

Then, the distribution of $[X_{k+1}^1 \ldots X_{k+1}^{W+S} \ldots X_K^1 \ldots X_K^{W+S}]^T$ with respect to the aggregated observation can be updated via the formulae in Appendix A. The updated distribution is denoted as follows:

$$f_{X_{k+1,\cdots,K}|X_{1,\cdots,k}}(x_{k+1,\cdots,K}^{1,\cdots,S+W}/x_{1,\cdots,k}^{1,\cdots,S+W}) = \sum_{l=1}^{M} \omega_l'(x_{1,\cdots,k}^{1,\cdots,S+W}, \Omega) \tag{24}$$
$$\times N_l(x_{k+1,\cdots,K}^{1,\cdots,S+W}; \mu_l(x_{1,\cdots,k}^{1,\cdots,S+W}, \Omega), \sigma_l(x_{1,\cdots,k}^{1,\cdots,S+W}, \Omega))$$

The updated distribution (24) has a GMM form.

*E. Integrating the updated distribution into the risk-limiting constraints*

When the $k$th observation is obtained, the conditional joint distribution of $[X_{k+1}^1 \ldots X_{k+1}^{W+S} \ldots X_K^1 \ldots X_K^{W+S}]^T$ shown in (24) is incorporated into the risk-limiting constraints (4)(5) as follows:

*a)* In (4), when $t=k+1$, the distribution of random variables $[X_{k+1}^1 \ldots X_{k+1}^{W+S}]^T$ should be extracted from the joint distribution of $[X_{k+1}^1 \ldots X_{k+1}^{W+S} \ldots X_K^1 \ldots X_K^{W+S}]^T$. Note that (24) is a GMM, and random variables $[X_{k+1}^1 \ldots X_{k+1}^{W+S}]^T$ can be regarded as a linear transformation of $[X_{k+1}^1 \ldots X_{k+1}^{W+S} \ldots X_K^1 \ldots X_K^{W+S}]^T$:

$$\begin{bmatrix} X_{k+1}^1 \\ \vdots \\ X_{k+1}^{W+S} \end{bmatrix} = \underbrace{\begin{bmatrix} 1 & 0 \end{bmatrix}}_{E} \begin{bmatrix} X_{k+1}^1 \\ \vdots \\ X_{k+1}^{W+S} \\ \vdots \\ X_K^1 \\ \vdots \\ X_K^{W+S} \end{bmatrix} \tag{25}$$

Using g *Lemma 1* in Appendix B, one can obtain a distribution of $[X_{k+1}^1 \ldots X_{k+1}^{W+S}]^T$:

$$f_{X_{k+1}|X_{1,\cdots,k}}(x_{k+1}^{1,\cdots,S+W}/x_{1,\cdots,k}^{1,\cdots,S+W}) = \sum_{l=1}^{M} \omega_l'(x_{1,\cdots,k}^{1,\cdots,S+W}, \Omega) \times \tag{26}$$
$$N_l(x_{k+1}^{1,\cdots,S+W}; E\mu_l(x_{1,\cdots,k}^{1,\cdots,S+W}, \Omega), E\sigma_l(x_{1,\cdots,k}^{1,\cdots,S+W}, \Omega)E^T)$$

Obviously, this is a GMM.

*b)* In (4), when $t>k+1$, the distribution of random variables $[X_{t+1}^1 \ldots X_{t+1}^{W+S}]^T$ should be conditioned on $[X_t^1 \ldots X_t^{W+S}]^T$. However, in the $k$th period, $[X_t^1 \ldots X_t^{W+S}]^T$ have not been observed. Similar to RLD, the idea of the proposed risk-limiting load restoration strategy is to utilize the conditional expectation of $[X_t^1 \ldots X_t^{W+S}]^T | [X_k^1 \ldots X_k^{W+S}]^T$ to replace $[X_t^1 \ldots X_t^{W+S}]^T$. The distribution of $[X_t^1 \ldots X_t^{W+S}]^T | [X_k^1 \ldots X_k^{W+S}]^T$ can be obtained in a similar way as (25)(26), being a GMM. Thereafter, one can use *Lemma 2* in Appendix C to compute the conditional expectation of $[X_t^1 \ldots X_t^{W+S}]^T | [X_k^1 \ldots X_k^{W+S}]^T$. Finally, using the formulae in Appendix A, one can obtain the conditional distribution of $[X_{t+1}^1 \ldots X_{t+1}^{W+S}]^T | [X_t^1 \ldots X_t^{W+S}]^T$. Due to the *conditional invariance* of GMM, the obtained distribution is still a GMM.

According to *a)* and *b)*, the distributions of $[X_t^1 \ldots X_t^{W+S}]^T$ for $t=k+1$ and $t>k+1$ are GMMs. For brevity, they are denoted as follows:

$$f_{X_t^{1,\cdots,S+W}|X_{1,\cdots,k}^{1,\cdots,S+W}}(x_t^{1,\cdots,S+W}/x_{1,\cdots,k}^{1,\cdots,S+W}) = \sum_{l=1}^{M} \omega_{l,t}'(x_{1,\cdots,k}^{1,\cdots,S+W}, \Omega)$$
$$\times N_{l,t}(x_t^{1,\cdots,S+W}; \mu_{l,t}(x_{1,\cdots,k}^{1,\cdots,S+W}, \Omega), \sigma_{l,t}(x_{1,\cdots,k}^{1,\cdots,S+W}, \Omega)) \tag{27}$$
$$\forall t \geq k+1$$

*c)* Since the risk-limiting constraint (5) is associated with random variables $[X_{k+1}^1 \ldots X_{k+1}^{W+S} \ldots X_K^1 \ldots X_K^{W+S}]^T$, the joint distribution in (24) is used for the risk-limiting constraint (5) without modifications.

*F. Comparison with a probabilistic forecasting tool*

Usually, a probabilistic forecasting tool is designed to obtain quantitative information of renewable generation random variables conditioned on explanatory variables [38]. Take the wind power as an example. The quantitative information could be confidence intervals for any quantile and/or full PDFs of wind generation. Typical explanatory variables include the forecasted wind speed from a NWP system as well as forecasted wind generation from a point forecasting tool. If one regards the past observations $[X_1^1 \ldots X_1^{W+S} \ldots X_t^1 \ldots X_t^{W+S}]^T$ as explanatory variables, the proposed method becomes a probabilistic forecasting tool for the future wind generation $[X_{t+1}^1 \ldots X_{t+1}^{W+S} \ldots X_K^1 \ldots X_K^{W+S}]^T$. Compared with the traditional probabilistic forecasting tools [38], the proposed method has the following features and advantages:

(1) The traditional methods rely on auxiliary forecasted information of explanatory variables. In contrast, the proposed method only relies on observations, which simplifies the implementation and makes the method appropriate in cases where the auxiliary forecasted information is not available.

(2) Usually, the traditional methods provide univariate PDFs and/or quantiles of random variables $X_{t+1}^1$, ..., $X_{t+1}^{W+S}$, ..., $X_K^1$, ..., $X_K^{W+S}$, respectively. If so, it is not easy to handle the arithmetical addition operations of random variables in the probabilistic constraints (4)(5), as $X_{t+1}^1$, ..., $X_{t+1}^{W+S}$, ..., $X_K^1$, ..., $X_K^{W+S}$ are correlated random variables. The proposed method obtains a joint distribution of a random vector $[X_{t+1}^1 \ldots X_{t+1}^{W+S} \ldots X_K^1 \ldots X_K^{W+S}]^T$. The joint distribution has a GMM form, which makes the computation of the probabilistic constraints (4)(5) feasible.

*G. Advantages*

The proposed method modeling uncertainties has the following advantages:

(1) Even if distributions of power outputs of WTs and PVs are different from one another, the proposed method is able to provide a universal model with satisfactory estimation by adjusting the parameter set $\Omega$. Also, the spatial-temporal correlation can be taken into account.

(2) Based on latest collected observations, the proposed method is able to infer distributions recursively as time rolls forward, increasing the accuracy of descriptions of future uncertainties.

(3) The updated distribution in each period is a GMM, facilitating the solution of the original problem (2)-(14). This point will be detailed in the next section.

## IV. SOLUTION METHODOLOGY

This section transforms probabilistic constraints (4)(5) into equivalent deterministic linear inequailities.

*A. Equivalent transformation of power constraint (4)*

Define the aggregation of $X_t^v$ in period $t$ by

$$X_t^{\text{sum}} = \sum_{v=1}^{W+S} X_t^v \qquad \forall t \geq \kappa+1 \quad (28)$$

Based *Lemma 1* in Appendix B, the distribution of $X_t^{\text{sum}}$ is obtained as follows,

$$f_{X_t^{\text{sum}}}(x_t^{\text{sum}}) = \sum_{l=1}^{M} \omega'_{l,t}(\mathbf{x}_{1,\cdots,k}^{1,\cdots,S+W}, \Omega) \times$$
$$N_l\left(x_t^{\text{sum}}; \mathbf{e}^T \boldsymbol{\mu}_{l,t}(\mathbf{x}_{1,\cdots,k}^{1,\cdots,S+W}, \Omega), \mathbf{e}^T \boldsymbol{\sigma}_{l,t}(\mathbf{x}_{1,\cdots,k}^{1,\cdots,S+W}, \Omega)\mathbf{e}\right) \quad (29)$$
$$\forall t \geq k+1$$

where **e** is a unit vector with a proper dimension.

Note that *Lemma 1* holds even if $X_t^v$ is correlated with others. Hence, the correlation of random variables can be handled.

Thereafter, moving the deterministic terms of (4) into left side and random variables to the right side, one can convert the probabilistic constraint (4) into (30) as follows:

$$\sum_{l=1}^{L} u_{t,l} P_{t,l} - \sum_{g=1}^{G} P_{t,g} \leq \text{CDF}_{X_t^{\text{sum}}}^{-1}(1-\alpha) \qquad \forall t \geq \kappa+1 \quad (30)$$

where CDF$^{-1}(\cdot)$ denotes the quantile of $X_t^{\text{sum}}$, which can be computed through its distribution function (29).

*B. Equivalent transformation of energy constraint (5)*

Adopting a similar idea as (28) and (29), one can compute the distribution of the aggregation of $X_t^v$ over multiple periods from $k+1$ to $T$ as follows:

$$X_{\kappa+1,T}^{\text{sum}} = \sum_{t=\kappa+1}^{T} \sum_{v=1}^{W+S} X_t^v \quad (31)$$

$$f_{X_{\kappa+1,T}^{\text{sum}}}\left(x_{\kappa+1,T}^{\text{sum}}\right) = \sum_{l=1}^{M} \omega'_l\left(\mathbf{x}_{1,\cdots,\kappa}^{1,\cdots,S+W}, \Omega\right) \times \quad (32)$$
$$N_l\left(x_{\kappa+1,T}^{\text{sum}}; \mathbf{e}^T \boldsymbol{\mu}_l\left(\mathbf{x}_{1,\cdots,\kappa}^{1,\cdots,S+W}, \Omega\right), \mathbf{e}^T \boldsymbol{\sigma}_l\left(\mathbf{x}_{1,\cdots,\kappa}^{1,\cdots,S+W}, \Omega\right)\mathbf{e}\right)$$

Thereafter, the equivalent transformation of (5) is:

$$\sum_{t=\kappa+1}^{T}\sum_{l=1}^{L} u_{t,l} P_{t,l} - \frac{1}{\tau}\sum_{g \notin ESS} EN_g(\kappa+1) \leq \text{CDF}_{X_{\kappa+1,T}^{\text{sum}}}^{-1}(1-\alpha) \quad (33)$$

where CDF$^{-1}(\cdot)$ of $X_{\kappa+1,T}^{\text{sum}}$ can be computed through its distribution function (32).

*C. Equivalent MILP*

At this point, the original problem in Section III is converted to an equivalent MILP as follows:

Objective:    Max (2)

Subject to:    (6)-(14), (30), and (33)

The MILP can be solved with available commercial solvers, such as *intlinprog* in MATLAB [42].

*D. Implementation procedure*

The implementation procedure of the proposed scheme for the $k$th period is shown in Fig. 7.

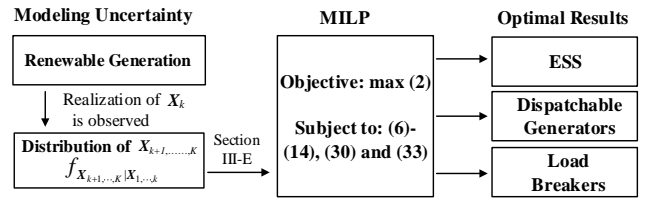

Fig. 7 Implementation of the proposed scheme

First, measurements of $[X_k^1 \ldots X_k^{W+S}]^T$ are obtained, based on which the joint distribution of future uncertainties is updated. Then, the joint distribution is fed into the risk-limiting constraints (4)(5), enabling the constraints to be converted into linear inequalities. Thereafter, MILP is solved. Finally, the optimal results, i.e., loads to be restored, the scheduled power of dispatchable generators, and the charging/discharging strategy of ESSs are sent to equipment for implementation.

*E. Optimal control dynamic dispatch (OCDD) formulation*

In the literature [43], the model (2)-(14) is categorized as "dynamic economic dispatch (DED)", because $P_{t,g}$ are decision variables. There is also another methodology for power system operation called "optimal control dynamic dispatch", in which the ramp rate $r_{t,g}$ are decision variables [43].

These two models are subject to similar constraints and implemented periodically. However, authors in [43] indicate that there are two differences between the two models:

(1) The optimal decision of the OCDD depends on initial values of $P_{0,g}$, while the DED does not.

(2) The OCDD considers the ramp limit on $P_{1,g}$ -$P_{0,g}$, while the DED neglects such a constraint.

The load restoration problem in this paper can be formulated as OCDD. The full OCDD formulation is provided in Appendix E. The proposed method to update distributions of uncertainties and the solution methodology to handle probabilistic constraints (4)(5) can be readily used in the OCDD formulation.

## V. CASE STUDY

### A. Test system and data

In the following tests, there are three microgrids, each with a diesel generator and an ESS. Besides, MG1 and MG2 both have a WT; MG3 has a PV. The duration of an outage from 07:00 to 17:00 is 10 hour with 1-hour time resolution. As a result, the total number of random variables is 30 (=3×10). Hourly historical data of WTs/PVs comes from publicly available datasets, i.e., "*solar integration data set*" and "*wind integration data set*", of NREL [36]. The data set consists of actual power outputs of wind/solar generation for 3 years. This paper uses data records of the first 1000 days as the "training set", and the rest as "test set". The "training set" is used to estimate the parameters of GMM using the maximum likelihood estimation technique, while the "test set" is used to test the performance of the proposed method. The proposed method is applicable to other time resolution of data, e.g., 30 min, 10 min, if needed. Load information of the IEEE 342-node system comes from Pacific Northwest National Laboratory (PNNL) [44]. Important information of the test system is listed in Table I.

TABLE I
IMPORTANT SYSTEM INFORMATION

| MG No. | Information | |
|---|---|---|
| MG1 | Diesel1 | : 2.0 MW, 5.0 MWh |
| | ESS1 | : 0.5 MW, 2.0 MWh, SOC70% |
| | WT1 | : 2.0 MW |
| | Loads | : 32 loads, 6.48MW in total |
| MG2 | Diesel2 | : 3.0 MW, 8.0 MWh |
| | ESS2 | : 1.5 MW, 3.0 MWh, SOC60% |
| | WT2 | : 2.0 MW |
| | Loads | : 30 loads, 8.79 MW in total |
| MG3 | Diesel3 | : 2.5 MW, 10.0 MWh |
| | ESS3 | : 1.0 MW, 4.0 MWh, SOC70% |
| | PV3 | : 2.0 MW |
| | Loads | : 34 loads, 6.54 MW in total |

### B. Validation of modeling uncertainties

This subsection provides test results using GMM to model uncertainties and updating distributions based on observations.

First, a GMM with 40 components is adopted to fit the historical data of actual power outputs of WTs/PVs over 10 periods. The results of WT1 are detailed in the following. Similar results can be obtained for WT2 and PV3.

To quantify the fitting performance, this paper uses three probabilistic metrics.

(1) The likelihood function value is used [41]. This is a positively-oriented score: the higher, the better. Note that the likelihood function value is not an "*absolute value metric*", which means there is not a universal threshold to distinguish between accuracy and inaccuracy. To overcome this hurdle, this paper compares the likelihood function of GMM with three widely-used statistical models [45]. They are the Gaussian model, Gaussian Copula model, and t Copula model. By doing so, the fitting performance of GMM is evaluated. The likelihood function values of GMM and the other three models are listed in Table II. It can be seen that GMM has the largest value, remarkably outperforming the other three.

TABLE II
FITTING TEST OF MODELING $[X^T\ Y^T]^T$

| Methods | Log-likelihood function values($10^3$) |
|---|---|
| Gaussian | 1.8345 |
| Gaussian Copula | 3.7658 |
| t Copula | 4.9025 |
| GMM | 7.1324 |

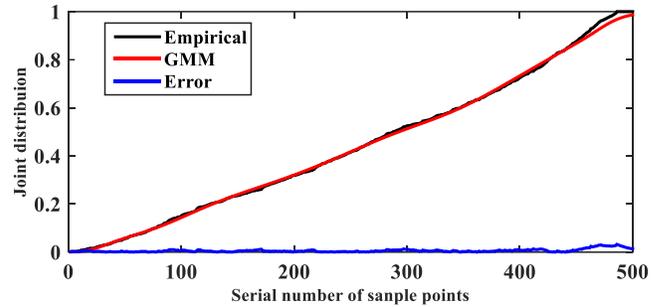

Fig. 8 Joint distributions of $X$ at 500 points

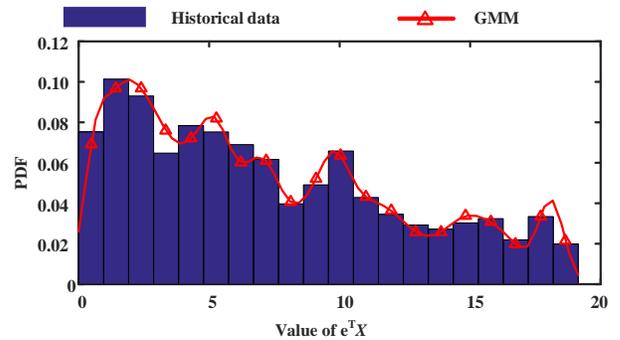

Fig. 9 GMM fits historical data of $\mathbf{e}^T X$

(2) The joint distribution of $X$ based on GMM is compared with the empirical joint distribution of $X$. According to *Glivenko–Cantelli theorem* [46], the empirical distribution converges to the true underlying distribution with probability one. Therefore, the empirical distribution can be used as a benchmark. The values of GMM based joint distribution and the empirical distribution at 500 points are shown in Fig. 8. The *maximum absolute error* (MAE) [47] and *root mean square error* (RMSE) [47] over the 500 points are computed. Both MAE and RMSE are small (0.0324, 0.0089, respectively). The results of Fig. 8 verify that GMM based joint distribution coincides with the empirical distribution, i.e., GMM well represents the historical data.

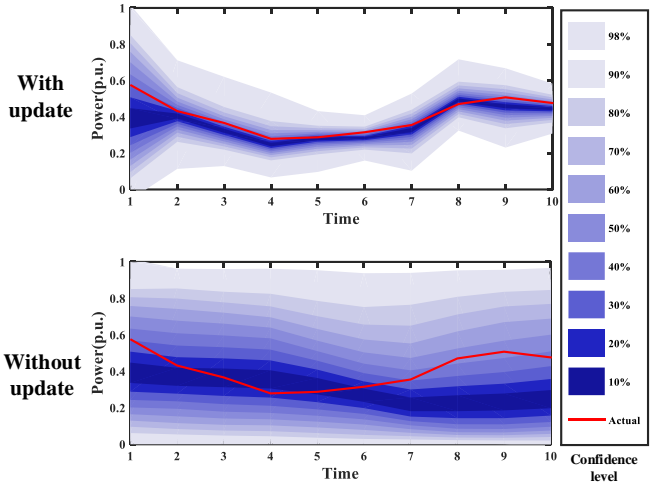

Fig. 10 distributions of WT1 in MG1 over 10 hours with/without update

(3) Note that when $X$ is represented by a GMM, the PDF of an aggregated random variables $\mathbf{e}^T X$ can be obtained based on *Lemma 1*. Obviously, if a GMM well represents $X$, the PDF of $\mathbf{e}^T X$, which is derived from GMM, should match well with the historical data of $\mathbf{e}^T X$. Based on this principle, this paper compares the PDF of $\mathbf{e}^T X$ derived from GMM with the histogram of $\mathbf{e}^T X$. The results are shown in Fig. 9. It can be seen that GMM matches well with the histogram. MAE and RMSE are also small (0.0112, and 0.0045, respectively). The results further indicate that GMM is a good representation for $X$.

Secondly, this paper computes the recursively updated distributions. In Fig. 10, WT1 in MG1 is taken as an example. The width of confidence levels of the updated distribution becomes narrower as time rolls forward. Meanwhile, the actual power is well bounded within confidence intervals. As a comparison, with the same confidence level, the width of the distribution without update is relatively broad, indicating a rough description of wind power uncertainties. The results reveal that the updated distribution provides a more elaborate description of wind power uncertainties than the one without updating. Similar results are obtained for WT2 in MG2, and PV3 in MG3.

*C. Results of the load restoration plan*
*1) Individual microgrid case*

With the validated model of uncertainties, this subsection is concerned with the load restoration using the updated distributions. In this test, each microgrid operates in a stand-alone mode with no connection to others, instructs its own energy resources, and supplies its own loads. In order to show the advantage of the recursively updated distribution, the distribution without update is provided as a comparison.

The load restoration strategy is tested with the "test set" in an operation simulation. First, a generation/load instruction for the next arriving time is obtained. At this point, restored loads are determined. Then, as the next period arrives, power outputs of WTs/PVs are realized. Since power outputs of WTs/PVs are random, an adjustment is activated: if the power supply is less than the restored load demand, the dispatchable generators should increase outputs and consume resources to meet the load demand; otherwise, there should be wind and/or solar spillage. For brevity, it is assumed that there is no way to regulate power outputs of ESSs once they are determined. By doing so, it is straightforward and convenient to use the restored loads, generator regulations, and wind/solar spillage to evaluate the effectiveness of the load restoration strategy.

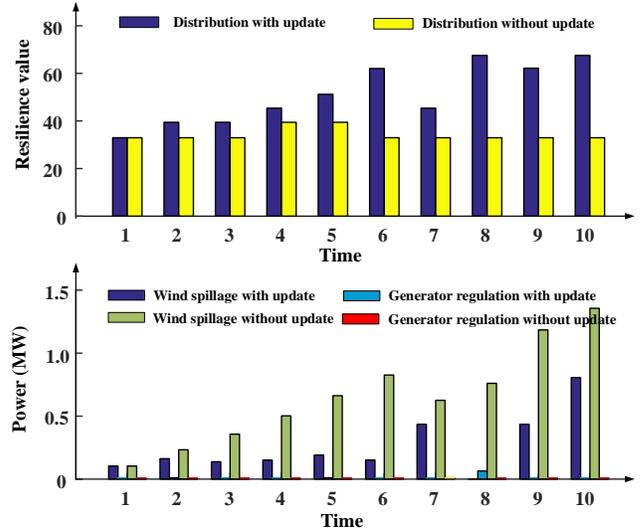

Fig. 11 Resilience values, generator regulations, and wind spillage of MG1

Fig. 11 shows resilience values, wind spillage, and generator regulations of MG1 (MG2 and MG3 are omitted). Three results can be obtained from Fig. 11:

(1) At the beginning, the resilience values of distributions with/without update are the same, since there has not been any observation to update the distribution.

TABLE III
COMPARISON OF DISTRIBUTION WITH/WITHOUT UPDATE

| Case | Resilience values |
|---|---|
| Distribution with update | 513.3 |
| Distribution without update | 342.6 |

(2) As time rolls forward, the recursively updated distribution takes advantages with more restored loads and less wind spillage over the distribution without update. This can be explained as follows: based on the distribution without update which provides a rough description of wind uncertainties, operators have to make conservative decisions to ensure that the probabilistic power/energy adequacy constraints (4)(5) are satisfied. That is, one uses more energy from the diesel generator of limited generation resources to serve loads, resulting in more wind spillage. As a comparison, the updated distribution provides an accurate estimation for the future uncertainties. Therefore, even with more wind power utilized, operators are confident that the adequacy requirements (4)(5) can be satisfied. The resilience values over the outage duration are listed in Table III. The results also verify that the updated distribution outperforms the distribution without update.

(3) An interesting phenomenon in Fig. 11 is that the wind spillage appears more frequently than the generator regulation. Here is an explanation: the confidence levels $\alpha$ in probabilistic constraints (4)(5) are 90%. That is, once the restored loads are determined, there is only a 10% chance that the load demand cannot be met, i.e., the generator regulation is thereafter activated. Therefore, in this test, the generator regulation is infrequently observed.

*2) Networked microgrid case*

In this test, the three microgrids connect with each other, become networked microgrids, and share resources to restore loads. This subsection discusses the advantage of networked microgrids over stand-alone ones. The resilience values of networked microgrids are computed. In Fig. 12, the networked resilience values are greater than the summation of three resilience values in stand-alone microgrids. This is reasonable: in networked microgrids, generation resources from one microgrid can be transferred to another microgrid to restore critical loads if needed.

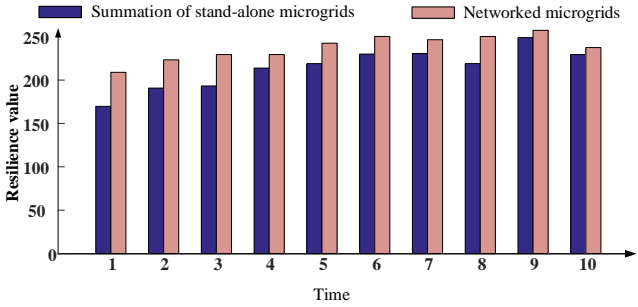

Fig. 12 Comparison of restored loads between networked microgrids and stand-alone microgrids

### D. Computation time

As far as the computation time is concerned, all tests are implemented on a Core-i5 PC, with a 2.39-GHz processor and 8 GB of RAM. The coding environment is Matlab. In the rolling plan, the optimization problem at the first period ($k$=0) has the heaviest computational burden since it determines decisions with the longest optimization window (see Fig. 3). It takes 11.649s to solve the problem for the individual case and 34.889s for the networked case. Note that the time resolution of the problem is set to be 1 hour in this paper. Therefore, the proposed method meets the time requirement. Even if the time resolution is reduced to 10 min, the proposed method is still applicable.

As far as communication latency is concerned, this paper assumes that the communication structure follows the standard IEC 61850 via TCP/IP [22]. Since a microgrid is not a large system, the communication delay is not significant. Suppose that the communication latency is 1 second. The computation and communication time together is within a minute. Therefore, the communication latency does not cause a major hurdle.

## VI. CONCLUSION

To enhance the resilience of a distribution system with microgrids, this paper proposes a method using observations to infer distributions of uncertainties of WTs/PVs based on the *conditional invariance* property of a GMM. The transformation of the original risk-limiting load restoration problem into an MILP benefits from the recursively updated distributions being a GMM.

This paper is focused on uncertainties. The power/energy adequacy requirements are considered, laying a foundation for load restoration with intermittent energy resources for practical use. Furthermore, concerning the operational and dynamic constraints, there are two on-going studies:

(1) Transmission limits should be considered. If the power flow is linear, e.g., the linearized DistFlow, a method proposed in previous authors' work [48] is readily applicable to deal with linear chance-constrained line power limits without major change. For the AC power flow in which the line power is a nonlinear implicit function of random power injections, the tractable computation of the risks pertaining to line limits remains an issue to resolve. The sample average approximation (SAA) is a potential solution.

(2) Dynamic constraints should be considered. Usually, distributed generators and ESSs have limited capabilities to withstand large transient shocks. Therefore, it is necessary to run dynamic simulations to ensure that the load restoration strategy does not cause instability or damage to equipment. Further discussions that are useful for the future work can be found in [6], [7].

## APPENDIX A
### CONDITIONAL INVARIANCE OF A GMM

Let the random vector $X$ be decoupled into two parts:
$$X := \begin{bmatrix} Y^T & Z^T \end{bmatrix}^T \quad (34)$$
where $Y$ represent observations, and $Z$ represent the rest entries of $X$.

The distribution of $X$ is a GMM. That is:
$$f_X(x) = f_{YZ}(y,z) = \sum_{m=1}^{M} \omega_m N_m(y,z \ ; \ \mu_m, \sigma_m) \quad (35)$$

$$\mu_m := \begin{bmatrix} \mu_m^y \\ \mu_m^z \end{bmatrix}, \sigma_m = \begin{bmatrix} \sigma_m^{yy} & \sigma_m^{yz} \\ \sigma_m^{zy} & \sigma_m^{zz} \end{bmatrix} \quad (36)$$

Then, the conditional distribution of $Z$ with respect to $Y=y$ is given as follows [49]:
$$f_{Z|Y}(z|y) = \sum_{l=1}^{M} \omega'_l(y,\Omega)_l \, N_l(z \ ; \mu_l(y,\Omega), \sigma_l(y,\Omega)) \quad (37)$$

$$\omega'_l(y,\Omega) = \omega_l \frac{N_l(y \ ; \ \mu_l^y, \sigma_l^y)}{\sum_{m=1}^{M} \omega_m N_m(y \ ; \ \mu_m^y, \sigma_m^y)} \quad (38)$$

$$\mu_l(y,\Omega) = \mu_l^z + \sigma_l^{zy}(\sigma_l^{yy})^{-1}(y - \mu_l^y) \quad (39)$$

$$\sigma_l(y,\Omega) = \sigma_l^{zz} - \sigma_l^{zy}(\sigma_l^{yy})^{-1}\sigma_l^{yz} \quad (40)$$

The distribution of $Z$ with respect to $Y=y$ is a GMM, enabling a recursive way to update distributions.

## APPENDIX B
### LINEAR INVARIANCE OF A GMM

***Lemma 1 [49]*** : If the distribution of $X$ is a GMM as in (16), and $X_{\text{LT}}$ is a defined as a linear transformation of $X$:

$$X_{\text{LT}} = \mathbf{A}X + \mathbf{C} \quad (41)$$

Then, the distribution of $X_{\text{LT}}$ is computed as follows:

$$f_{X_{\text{LT}}}(x_{\text{LT}}) = \sum_{m=1}^{M} \omega_m N_m\left(x_{\text{LT}}; \ \mathbf{A}\boldsymbol{\mu}_m + \mathbf{C}, \mathbf{A}\boldsymbol{\sigma}_m \mathbf{A}^{\text{T}}\right) \quad (42)$$

*Lemma 1* holds even if entries of $X$ are correlated.

## APPENDIX C
### EXPECTATION OF A GMM

***Lemma 2***: If the distribution of $X$ is a GMM as in (16), the expectation of $X$ is

$$\mathbf{E}[\mathbf{X}] = \sum_{m=1}^{M} \omega_m \boldsymbol{\mu}_m \quad (43)$$

## APPENDIX D
### COMPARISON WITH PERSISTENCE FORECASTING

#### A. Modification of probabilistic constraints (4)(5)

Since "persistence forecasting" is a point forecasting tool of uncertainties, the probabilistic constraints (4)(5) should be modified as follows:

$$\sum_{g=1}^{G} P_{t,g} + \sum_{v=1}^{W+S} X_t^v(\text{fore}) \geq \sum_{l=1}^{L} u_{t,l} P_{t,l} \quad \forall t \geq \kappa + 1 \quad (44)$$

$$\sum_{g \notin ESS} EN_g(\kappa+1) + \tau \sum_{t=\kappa+1}^{T} \sum_{v=1}^{W+S} X_t^v(\text{fore}) \geq \sum_{t=\kappa+1}^{T} \sum_{l=1}^{L} P_{t,l} \tau \quad (45)$$

where $X_t^v(\text{fore})$ represents the forecasted value from "persistence forecasting" technique.

If one wants to compare the proposed method with "persistence forecasting" technique, the *updated distribution* should be modified as a point prediction. An intuitive idea is using the expectation of the *updated distribution* as a point prediction. Thereafter, probabilistic constraints (4)(5) should be formulated as expectations. That is:

$$\sum_{g=1}^{G} P_{t,g} + \mathbf{E}\left[\sum_{v=1}^{W+S} X_t^v\right] \geq \sum_{l=1}^{L} u_{t,l} P_{t,l} \quad \forall t \geq \kappa + 1 \quad (46)$$

$$\sum_{g \notin ESS} EN_g(\kappa+1) + \mathbf{E}\left[\sum_{t=\kappa+1}^{T} \sum_{v=1}^{W+S} X_t^v \tau\right] \geq \sum_{t=\kappa+1}^{T} \sum_{l=1}^{L} P_{t,l} \quad (47)$$

where $\mathbf{E}[\cdot]$ is the expectation operation.

The distributions of the aggregation of $X_t^v$ have been derived in (29)(32), which are GMMs. Therefore, the expectations can be computed based on *Lemma 2* in Appendix C.

At this point, one can compare "persistence forecasting" with the proposed *updated expectation*.

#### B. Test results

The test procedure is the same as those in Section V-C. The results are shown in Fig. D-1 and Table D-I. It can be seen that the proposed method with *updated expectations* obtains a higher resilience value, surpassing the "persistence forecasting" method.

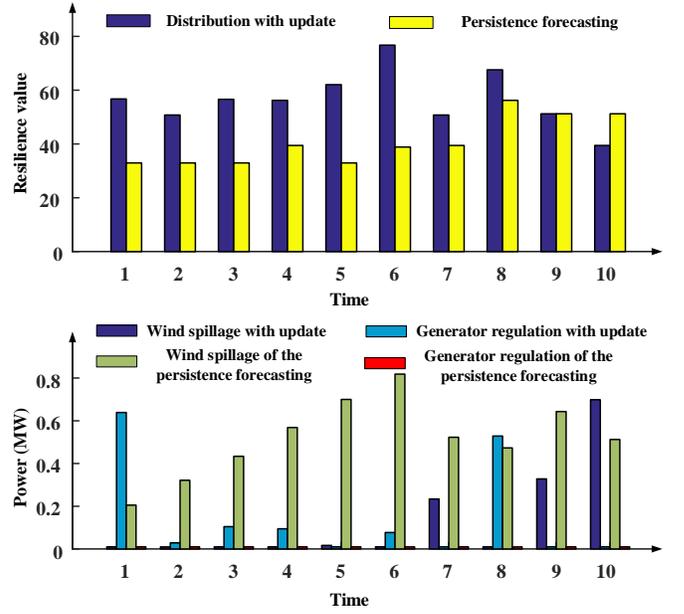

Fig. D-1 Resilience value, generator regulations, and wind spillage of MG1

TABLE D-I
RESILIENCE VALUE COMPARISON

| Case | Resilience value |
| --- | --- |
| Proposed updated expectation | 568.2 |
| Distribution without update | 408.3 |

## APPENDIX E
### AN EXTENSION TO THE OCDD MODEL

#### C. Full formulation

● Risk limiting constraints:

$$\Pr\left\{\sum_{g=1}^{G}\left(P_{0,g} + \sum_{i=1}^{t} r_{i,g}\right) + \sum_{v=1}^{W+S} X_t^v \geq \sum_{l=1}^{L} u_{t,l} P_{t,l}\right\} \geq \alpha \quad (48)$$

$$\Pr\left\{\begin{array}{c}\sum_{g \notin ESS} EN_g(\kappa+1) + \sum_{t=\kappa+1}^{T} \sum_{v=1}^{W+S} X_t^v \tau \\ \geq \sum_{t=\kappa+1}^{T} \sum_{l=1}^{L} u_{t,l} P_{t,l} \tau\end{array}\right\} \geq \alpha \quad (49)$$

● Deterministic constraints

$$P_g^{\min} \leq P_{0,g} + \sum_{i=1}^{t} r_{i,g} \leq P_g^{\max} \quad (50)$$
$$\forall t \geq \kappa + 1, \ \forall g \notin ESS$$

$$\sum_{t=\kappa+1}^{T}\left(P_{0,g} + \sum_{i=1}^{t} r_{i,g}\right) \leq EN_g(\kappa+1) \quad (51)$$
$$\forall g \notin ESS$$

$$\chi_{t,g} + \gamma_{t,g} \leq 1 \quad \forall t \geq \kappa + 1, \ \forall g \in ESS \quad (52)$$

$$0 \leq P_{0,g}^{\text{dch}} + \sum_{i=1}^{t} r_{i,g}^{\text{dch}} \leq \chi_{t,g} P_g^{\text{dch,max}} \quad (53)$$
$$\forall t \geq \kappa + 1, \ g \in ESS$$

$$-\gamma_{t,g} P_g^{\text{ch,max}} \leq P_{0,g}^{\text{ch}} + \sum_{i=1}^{t} r_{i,g}^{\text{ch}} \leq 0 \quad (54)$$
$$\forall t \geq \kappa + 1, \ g \in ESS$$

$$SOC_g^{\min} \leq SOC_{0,g} + \sum_{i=1}^{t} rsoc_{i,g} \leq SOC_g^{\max} \quad (55)$$
$$\forall t \geq \kappa+1, \ g \in ESS$$

$$rsoc_{t,g} = -\tau \left( P_{0,g}^{\text{dch}} + \sum_{i=1}^{t-1} r_{i,g}^{\text{dch}} \rho_d^{-1} + P_{0,g}^{\text{ch}} + \sum_{i=1}^{t-1} r_{i,g}^{\text{ch}} \rho_c \right) / EC_g \quad (56)$$
$$\forall t \geq \kappa+2, \ g \in ESS$$

*D. Test results*

This paper takes the optimization problem of the first period ($k$=0) as an example to show the difference between DED and OCDD models. As discussed in Section IV-E, the OCDD considers an initial value of $P_{0,g}$ and the ramp limit on $P_{1,g}$ -$P_{0,g}$, while the DED does not. To clearly illustrate this phenomenon, $P_{0,g}$ in OCDD is set to be 0. That is to say, the maximum value of $P_{1,g}$ equals to the ramping rate of generator $g$. In this paper, the ramping rate is set to be 0.3MW/h.

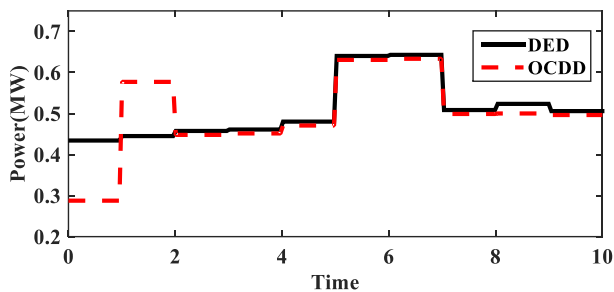
Fig. E-1 Scheduled power of the diesel generator in MG1

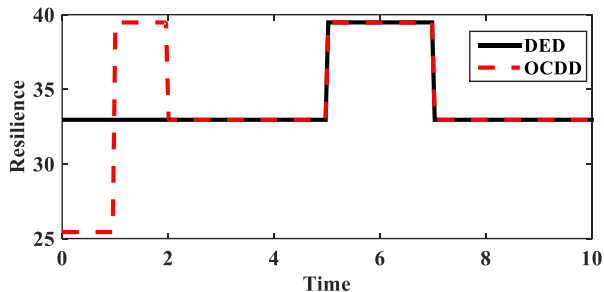
Fig. E-2 Objective functions of the two models

Tests are conducted on MG1. Results of the two models are shown in Fig. E-1 and Fig. E-2. In Fig. E-1, it is observed that the scheduled power of the diesel in OCDD model and that in DED model are different at the beginning, but they coincide later. Similar conclusions are obtained for the objective functions in Fig. E-2. Because the OCDD model has more constraints than DED, the optimal solution of OCDD is no better than that of DED. According to Fig. E-2, the objective functions of the two models are close to each other (342.57, 341.58, respectively). The results of the two figures are consistent with those reported in [43].